# The Impact of the Coronavirus Pandemic on New York City Real Estate: First Evidence

Jeffrey P. Cohen[*]  
University of Connecticut

Felix L. Friedt[†]  
Macalester College

Jackson P. Lautier[‡]  
University of Connecticut

**Abstract:** We investigate whether pandemic-induced contagion disamenities and income effects arising due to COVID-related unemployment adversely affected real estate prices of one- or two-family owner-occupied properties across New York City (NYC). First, OLS hedonic results indicate that greater COVID case numbers are concentrated in neighborhoods with lower-valued properties. Second, we use a repeat-sales approach for the period 2003 to 2020, and we find that both the possibility of contagion and pandemic-induced income effects adversely impacted home sale prices. Estimates suggest sale prices fell by roughly $60,000 or around 8% in response to both of the following: 1,000 additional infections per 100,000 residents; and a 10-percentage point increase in unemployment in a given Modified Zip Code Tabulation Area (MODZCTA). These price effects were more pronounced during the second wave of infections. Based on cumulative MODZCTA infection rates through 2020, the estimated COVID-19 price discount ranged from approximately 1% to 50% in the most affected neighborhoods, and averaged 14%. The *contagion* effect intensified in the more affluent, but less densely populated NYC neighborhoods, while the *income* effect was more pronounced in the most densely populated neighborhoods with more rental properties and greater population shares of foreign-born residents. This disparity implies the pandemic may have been correlated with a wider gap in housing wealth in NYC between homeowners in lower-priced and higher-priced neighborhoods.

[*] Jeffrey P. Cohen, Kinnard Scholar in Real Estate and Professor of Finance, Department of Finance and Center for Real Estate and Urban Economic Studies, School of Business, University of Connecticut, 2100 Hillside Rd, Storrs, CT 06269; and Research Fellow, Institute for Economic Equity, Federal Reserve Bank of St. Louis. jeffrey.cohen@business.uconn.edu

[†] Felix L. Friedt, Assistant Professor of Economics, Department of Economics, Macalester College, 1600 Grand Avenue, Saint Paul, MN 55105; ffriedt@macalester.edu.

[‡] Jackson Lautier, Department of Statistics, University of Connecticut, Philip E. Austin Building, 215 Glenbrook Road, Storrs, CT 06269-4120, jackson.lautier@uconn.edu.

**Acknowledgements:** Cohen's work on this paper was completed in part while he has been a Research Fellow at the Institute for Economic Equity, Federal Reserve Bank of St. Louis. Lautier's efforts on this paper were based upon work supported in part by the National Science Foundation Graduate Research Fellowship under Grant No. DHE 1747453. The authors would also like to thank Jason Barr for his helpful suggestions on the availability of related New York City datasets. Session participants at the 2022 ASSA/AREUEA Annual Meetings (virtual) provided helpful comments. Any remaining errors are our own. The views expressed in this paper are those of the authors and do not necessarily represent the Federal Reserve Bank of St. Louis, the Federal Reserve Board, or the Federal Reserve System. Revised 1/27/2022.

# 1 Introduction

The novel Coronavirus has transformed life as previously known throughout the world, and the changes in U.S. residential real estate markets are unprecedented. In the early months of 2020 during the pandemic, New York City (NYC)[1] experienced a disproportionate intensity of cases and deaths, relative to the rest of the United States, partly due to the relatively high density in many parts of the City. As the pandemic worsened between March 1, 2020 to July 31, 2020 (which we refer to as the first wave) and again surged between August 1, 2020 through December 31, 2020 (described herein as the second wave), vast numbers of NYC residents chose to leave the city for the suburbs, where more spacious residences and automobile-dependent towns may have been appealing due to their relatively insulated way of life. Although this major shift is still unfolding to some degree, it leads one to wonder how the novel Coronavirus is related to residential real estate prices in NYC. An abundance of detailed NYC COVID-19 case data with relatively precise locations of these cases within the City, lends to the ability to examine the strength of the relationship between the novel Coronavirus and residential real estate prices in the five boroughs. It also raises the question of whether homes in the City that are already located on larger lots of land (i.e., much higher priced) may not experience changes in the same ways as homes that are much closer together in some of its lower-priced neighborhoods. These are the issues that we study in this paper. Specifically, we investigate whether a pandemic-induced contagion disamenity adversely affected real estate prices of one- or two-family owner-occupied properties across NYC.

As the world is still reeling from subsequent waves of the virus, the impacts on real estate markets are an ongoing issue and the nascent literature in this area is still developing. The limited existing research on residential real estate includes analyses of U.S. house price responses to shutdowns and re-openings related to COVID-19, along with aggregate U.S. market impacts from COVID-19.

One of the existing U.S. studies is D'Lima et al. (2020), who consider the effect of closures and re-openings. Employing a difference-in-differences identification strategy, the authors assess the impacts of COVID-19 on residential real estate prices throughout the United States. Their approach follows the commonly used hedonic house price methods, which assume that the demand for housing can be broken out into an array of property characteristics (Rosen, 1974), such as numbers of bedrooms, bathrooms, square footage, acreage, etc., and amenities and/or disamenities (Banzhaf and McCormick, 2012). Then linear regression estimation of this hedonic price equation, where the natural logarithm of house price is the dependent variable, can yield elasticities of house prices with respect to the amenities/disamenities. D'Lima et al. (2020) include an indicator in their hedonic model for whether or not each property sale at time *t* was in a state that had shut down at time *t*, and a separate indicator for whether a property that sold at time *t* was in a state that had reopened at time *t*. These authors control for zip-code fixed-effects and cluster at the zip code level. Their findings include that the shutdowns had a statistically insignificant negative effect on prices, while reopening had a statistically significant negative effect (with elasticity of approximately 0.9). They argue that not all restrictions were immediately lifted upon reopening, which led to the negative effect of reopening.

---

[1] We also abbreviate NYC as 'the City' in some places for ease of exposition.

Zhao (2020) studies aggregate and micro (zip code level) effects of COVID-19 on U.S. housing markets. Aggregate level findings include that prices have risen much faster than they did in the months prior to the Global Financial Crisis; and there has been a structural break in housing demand after mortgage rates have fallen post-March 2020. Zhao (2020) also notes on the micro-level that individuals on the lowest and highest levels of the income distribution have demonstrated the greatest increases in housing demand. Finally, they note that changes in demand in cities are comparable to changes in demand in suburbs and rural areas.

Studies of the real estate market impacts of COVID-19 outside the U.S. are sparse. While Francke and Korevaar (2020), for example, examine how historical pandemics impact house prices in Amsterdam, they do not specifically consider COVID-19. Rather, they focus on other past pandemics. They find that there is a significantly negative effect within 6 months after a pandemic, but this discount is only temporary. This transitory finding potentially bodes well for the impacts of COVID-19 in many cities throughout the world. In another recent study beyond the U.S. context, the impact of deaths from COVID-19 on Chinese housing markets is the focus of Chong and Liu (2020). Their finding is that months with the fewest numbers of deaths experienced the lowest house price changes in response to more deaths. This relationship is reversed in months with larger numbers of deaths. The authors refer to this phenomenon as a "U-shaped" effect.

This review of the existing literature on residential real estate and COVID-19 indicates gaps that deserve further examination. For instance, a micro analysis of one particular city, and how local rates of COVID-19 cases are correlated with house prices in the very short term, is of interest. NYC is an excellent laboratory for this type of analysis, given the heterogeneity across its 5 boroughs and many hospitals in the city. Another desirable feature of our NYC focus is the recent migration of residents to the suburbs, at least in part as a way to try and evade potential future waves of the pandemic. By focusing on a specific 'big-city' real estate market and estimating the response to the intensity of the outbreak (i.e., cumulative case numbers) rather than shutdowns, we are able to assess heterogeneity in the estimated effects. That is, all of NYC was shut down for some time, and yet still there were differences in how the market changed across boroughs and/or Modified Zip Codes Tabulation Areas (MODZCTAs) due to variation in actual COVID-19 intensities and pandemic-induced surges in unemployment rates.

We have two separate prongs to our approach to estimate what we call the *contagion*[2] effect and *income* effect on NYC real estate markets. The first is a classic hedonic housing price model for NYC non-investor-owned one- or two-family house sales, where the dependent variable is house price at time *t*; we include control variables for the cumulative number of novel Coronavirus infections per 100,000 residents within the local MODZCTA, at time *t* ($CC_{it}$); as well as local MODZCTA unemployment rates ($U_{it}$). Across neighborhoods, we find a negative relationship between house prices and $CC_{it}$ as well as $U_{it}$, suggesting that NYC neighborhoods with more intense COVID-19 outbreaks and greater rates of unemployment experienced sale price discounts. This is the contagion effect. While these correlations are insightful, they may not reveal the causal impact of COVID-19 on NYC property prices. One potential issue may be changes to the neighborhood-specific composition of sold properties along some unobservable home attribute. One possible scenario supported by some anecdotal evidence (Goodman and Rashbaum, 2020; Haynes, 2020) is, for example, an exodus of some of the more affluent residents, such that COVID-19 triggers the sale of more valuable properties within NYC neighborhoods. As a result of this compositional shift, the observed average sale price in a given neighborhood may rise and offset the otherwise adverse impact of the pandemic on the value of all properties - sold and unsold.

In the second approach, we attempt to reconcile these initial findings of the underlying relationship by restricting ourselves to transactions of repeat-sales properties between 2003 and 2020. We differentiate the effects across the first and second waves of infections in 2020 and estimate a differenced model, where the difference in prices (between repeat sales) is regressed against the difference in cumulative case rates, changes in local unemployment, variations in investor ownership shares as well as fixed effects. This approach identifies the COVID-19 price effect by estimating the average change in sale prices at the property level and is therefore immune to the potential pandemic-induced compositional changes of real estate markets at the neighborhood level. We difference out time-invariant observable and unobservable home and neighborhood characteristics, such as whether a home is a high- or low-valued property in a low- or high-valued neighborhood, while controlling for changing local labor market (unemployment) and real estate conditions. As one way to address the potential issue of changes in the quality of the houses over time that are differenced out by the assumption that the characteristics are time-invariant, we also include an indicator for whether there have been modifications/renovations to the property. In addition to including the renovations indicator, using differences of the unemployment variable also addresses the common time-invariance critique of the repeat-sales approach. We find that the coefficient on the change in cumulative infection rates is negative and statistically significant in the repeat-sales models. In other words, as there are more cases per resident over time, this adversely affects house prices. This *contagion* effect on prices is more pronounced during the second wave of infections, when for every 1,000 additional COVID-19 infections per 100,000 residents in a given MODZCTA the value of a one- or two-family NYC home is found to decline by over $60,000 between two sales of the same property, on average. Our heterogeneity analysis suggests that buyers of higher priced homes are more risk averse with respect to rising case rates, while residents in already densely populated neighborhoods exhibit less reaction to contagion – or perhaps cannot afford to.

---

[2] We make the assumption that information on detailed location COVID-19 intensity was widely available, particularly for NYC residents. One example is the "NYC Coronavirus Case Count" interactive web hub made freely available by the *New York Times* and updated through May 21, 2021. See https://www.nytimes.com/interactive/2020/nyregion/new-york-city-coronavirus-cases.html.

In terms of the *income* effect, we find that a rise in unemployment rates significantly reduces home sale prices for repeat sales properties. Again, this is particularly apparent during the second wave of the pandemic. In comparison to pre-pandemic levels, NYC unemployment rates rose by more than 8 percentage points suggesting an average decline in home values of around $50,000. Based on the heterogeneity analysis, it appears that, while home buyers in more densely populated neighborhoods with greater dependence on public transit and higher rental shares are less sensitive to rising infection rates, the *income* effect on home prices significantly intensifies in these types of neighborhoods.

Interestingly, homes that sold for a price in the bottom quartile of the pre-COVID local price distribution experienced a significantly higher price discount in response to higher cumulative cases during the initial wave of the pandemic, but negligible price discount during the second half (or second wave) of 2020. In contrast, homes valued in the top quartile of the pre-COVID local price distribution experienced an intensifying price discount in response to rising cases.

These findings may be due in part to the fact that in NYC, homes on the lower end of the price distribution imply less wealth associated with homeownership relative to homes on the higher end of the distribution. This may have reduced the flexibility of lower-priced homeowners to move to another house in the suburbs more quickly during a time when the suburban housing market was very hot with cash offers among the winners and multiple bids over asking prices. Perhaps these short-run market frictions - both in NYC and in the suburbs - can be a source of explanation for the negligible price discount for lower-priced NYC houses during the second wave, as higher priced homeowners in NYC were able to get out of their NYC homes and into suburban houses with more agility. This difference in responses among owners of higher-priced and lower-priced houses may arise through the mechanism of (i) lower valued properties tended to experience more COVID cases, and (ii) we observe a negative effect of cases on house prices. But despite there being a much larger impact of cases on prices for lower valued homes in the "first wave" of COVID in NYC, the "second wave" of COVID in NYC is accompanied by an insignificant effect of cases on prices for lower-valued homes. This may suggest that residents in lower-valued properties have come to treat pandemic risk as a new neighborhood constant, like in areas of high-crime. In other words, this first wave effect was so intense on lower-valued homes that residents perceived this as a permanent change so that future upticks in cases may have been already priced-in to second wave market transactions.

All-in-all, our findings highlight the adverse effect COVID-19 has had on the NYC residential real estate market through December, 2020. Over the span of just 10 months since the first reported cases, COVID-19 infections and the associated rise in unemployment reduced average house sale prices by around 20% during the second wave of infections. In comparison to the expected pre-COVID home value appreciation, the pandemic had eroded a value equivalent to a 5 to 6-year expected return on residential property investment in just ten months. Moreover, these average effects are not equally distributed across the NYC universe of one- or two-family homes, but concentrated among lower-valued neighborhoods that experience greater rates of infections and face higher unemployment rates.

The remainder of this paper proceeds as follows. In the next section, we provide some background on NYC residential real estate and COVID-19. In section 3, we describe the empirical models that we implement, followed by a detailed description of our data in section 4. We present the results of our estimations in section 5, after which we provide some discussion and conclusions along with suggestions for future work.

# 2 Background

It is a rare occurrence to see an Op-Ed in the *New York Times* by a certain famous New York-based standup comedian and sitcom star defending the future of NYC (Seinfeld, 2020) against hyperbolic claims that NYC is "dead forever" (Altucher, 2020). Then again, these are exceptional times. Major news outlets have told a story of large-scale outbound migration from NYC due to Coronavirus restrictions. Manhattan vacancy rates climbed to 14-year high levels (Haag, 2020b). Conversely, surrounding suburbs were struggling to absorb the demand of "fleeing" New Yorkers. A home in East Orange, N.J., for example, was listed at $285,000, hosted 97 showings, received 21 offers, and went under contract 21% over asking – all over a three-day period in July of 2020 (Haag, 2020a).

Figures 1a through 1f provide a first look at the NYC housing market before and after the outbreak of the novel Coronavirus and corroborate some of the anecdotal evidence reported in these recent news articles. In each figure, we plot the number of daily new COVID-19 cases and 7-day moving average of the number of daily property sales from October 1, 2019 to December 31, 2020 (solid blue and green lines) and contrast the latter against the moving average number of daily real estate transactions in the previous year (i.e., October 1, 2018 through December 31, 2019) (dashed grey line). Figure 1a presents these data for all of NYC, whereas Figures 1b through 1f illustrate the real estate and pandemic dynamics for each of the five NYC boroughs.

Each of the six graphs clearly shows that the initial outbreak of the novel Coronavirus in NYC coincides with notable disruption of the city's real estate market. While daily sales across all property types in the second half of 2019 and early 2020 tend to track the volume of transactions in the previous year, the two series diverge starting in March 2020 as the first wave of cases are reported in NYC. In aggregate, Figure 1a shows that the number of daily sales falls by around 50 transactions ($\approx$ -50%) per day during the first few months of the COVID-19 outbreak in NYC and that this response not only persisted, but actually worsened to around 80 lost sales per day through July 31, 2020. The Bronx, Brooklyn and Queens, for example, showed an immediate 50% to 70% decline in the volume of the real estate market one month after the outbreak relative to pre-pandemic data (see Figures 1b, 1c, and 1e). The initial responses of the real estate markets in Manhattan and Staten Island were also negative, but appeared to be more moderate and short-lived (see Figures 1d and 1f).

While the decline in real estate market volume was sharp, the data for the second half of 2020 illustrate that initial claims may have been exaggerated. The number of property sales steadily recovered between July and December of 2020. Staten Island shows the most rapid recovery among all of the five NYC boroughs (Figure 1f). In aggregate, the NYC real estate market seems to have recovered in terms of volume by October 2020. Interestingly, each of the Figures illustrates that the onset of the second wave of infections in late 2020 does not appear to have the same adverse effects as the initial outbreak.

Using the detailed information on real estate transactions and spatially disaggregated data on the number of new COVID-19 infections, we also show these patterns at the MODZCTA level. Specifically, we calculate average daily infections and the average number of real estate transactions for MODZCTAs across the distribution of average pre-pandemic (i.e., 2003-2019) house sale prices. Figures 2a through 2d illustrate the changes for the average MODZCTAs in each quartile of the pre-pandemic home value distribution.

For MODZCTAs with average sale prices below the $75^{th}$ percentile the patterns are similar including a sizable collapse in real estate transactions in March and April 2020 and a prolonged recovery through roughly September or October 2020. Within these MODZCTAs, daily infections start to rise around the same time and peak at approximately the same levels (i.e., around 80 new cases per MODZCTA on April $9^{th}$ and approximately 170-180 new cases per MODZCTA on April $26^{th}$). In contrast, the average MODZCTA in the top quartile of the pre-pandemic house sale price distribution seems to experience a more moderate disruption of the local real estate market and is subject to a less intense outbreak (i.e., around 30 new cases per MODZCTA on April $9^{th}$ and 115 new cases per MODZCTA on April $26^{th}$).[3] Finally, we note that, irrespective of the level of spatial disaggregation, pre-pandemic sales patterns are similar. That is, the volumes of real estate transactions before the onset of the initial COVID-19 outbreak in late 2019 (October-December) and early 2020 (January-February) are on similar trajectories compared to the number of transactions late 2018 and early 2019. The initial collapse clearly coincides with the rise of infections among NYC residents and does not predate the pandemic. This appears to be true in aggregate (Figure 1a), across boroughs (Figures 1b-1f) as well as MODZCTAs with high to low value properties (Figures 2a-2d).

---

[3] There are a few possible explanations for these patterns. For example, wealthier individuals may have the financial means to better isolate themselves, which reduces the risk of infection. Moreover, housing wealth may be disproportionately held among older residents who face greater COVID infection risks and are therefore reducing the exposure to potential infections. It is also possible, however, that the likelihood of getting tested inversely correlates with housing wealth.

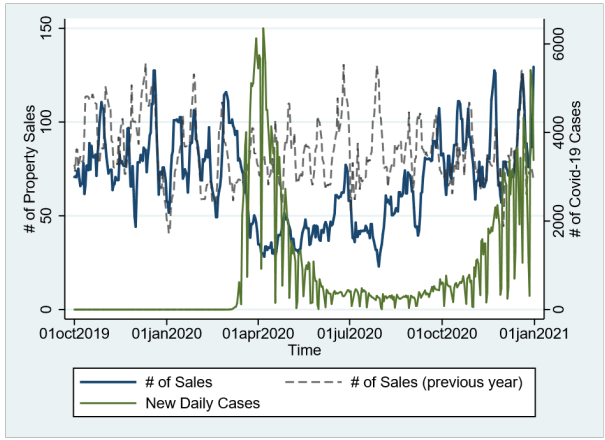
(a) New York City

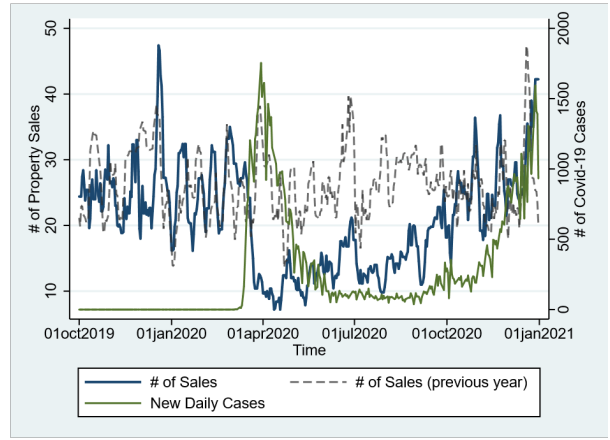
(b) Bronx

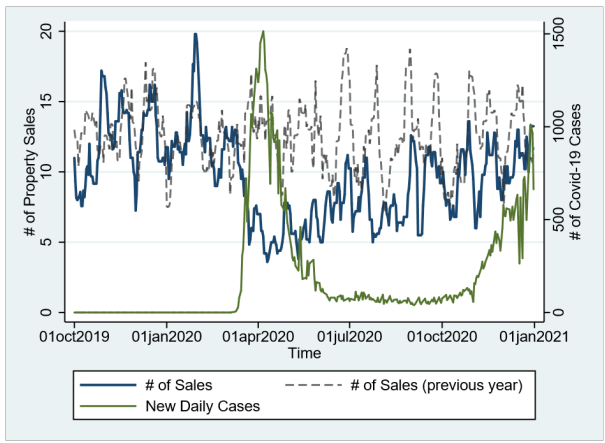
(c) Brooklyn

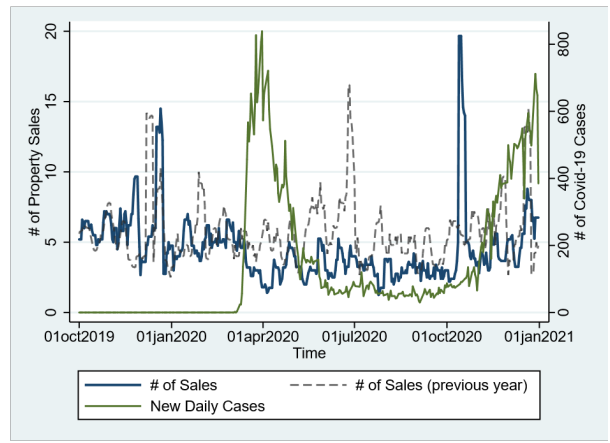
(d) Manhattan

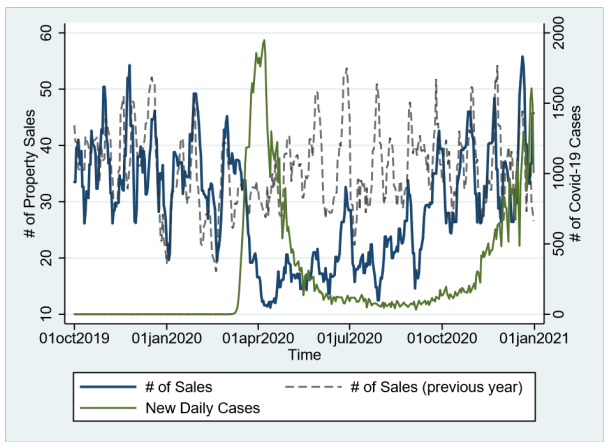
(e) Queens

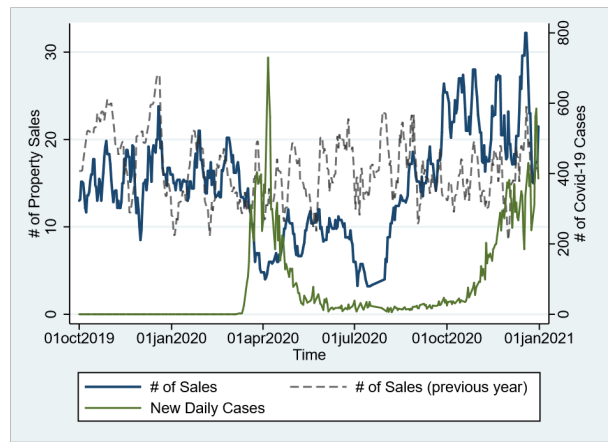
(f) Staten Island

Figure 1: NYC Daily Case Counts and Avg. # of House Sales by Borough

# 3  Model

To evaluate the impact of the COVID-19 pandemic on the NYC real estate market more formally, we first estimate the impact of novel Coronavirus infections on NYC property sale prices using the hedonic model popularized by Rosen (1974)[4] and used in many economic studies on real estate markets.[5] We are able to estimate the pandemic effects on the full sample of non-investor-owned one and two-family property transactions as well as the more restrictive repeat-sales sample. To begin, we derive the traditional hedonic model to evaluate the price effect of the COVID-19 pandemic on NYC homes. The classic hedonic model of Rosen (1974) has more recently been extended to incorporate amenities and disamenities. This hedonic model can be written, in general form, as follows:

$$P_{it} = f(H_i, AD). \tag{1}$$

In Equation (1), property prices are assumed to be a function of property characteristics, $H_i$, and amenities/disamenities (*AD*). Best practices for incorporating amenities/disamenities into hedonic models are described by (Bishop et al., 2020). In our specific application, the purpose of the hedonic model is to determine how various characteristics of the property and neighborhood demographics, as well as particular disamenities of interest (i.e., the rise in novel Coronavirus cases and associated rise in unemployment) affect property sale prices. Conceptually, the pandemic can influence the demand for homes through two primary channels including: 1) the adverse effects on *income*; and 2) the increased *contagion* in a densely populated city, such as NYC. We attempt to capture and differentiate these two mechanisms integrating COVID-19 case rates and local unemployment rates into the hedonic framework. By controlling for the demographics and characteristics with regression analysis, it is possible to estimate how additional infections and higher unemployment correlate with property sale prices. Therefore, our hedonic model takes the following form:

$$P_{it} = \beta_0 + \beta_1 CC_{bt} + \gamma H_i + \varphi E_{bt} + \omega N_{bt} + \alpha_b + \alpha_t + \varepsilon_{it}, \tag{2}$$

---

[4] Rosen (1974) popularized the study by Kain and Quigley (1970).
[5] For a review of this literature see, for example, Chau and Chin (2003), while an example of a specific application to another dis-amenity - aircraft noise pollution - is given by Cohen and Coughlin (2008).

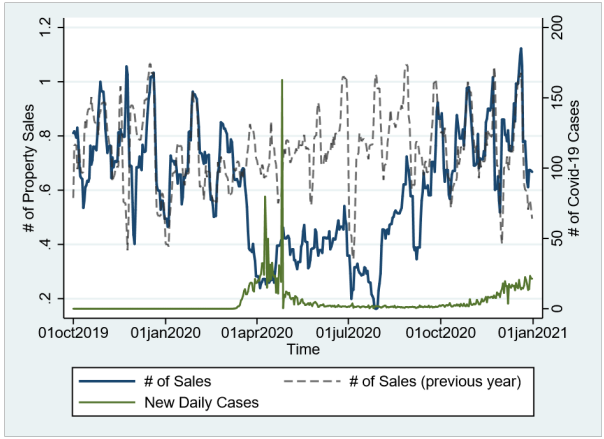
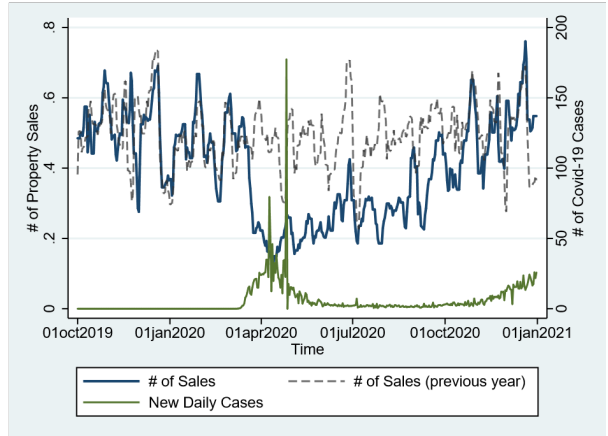
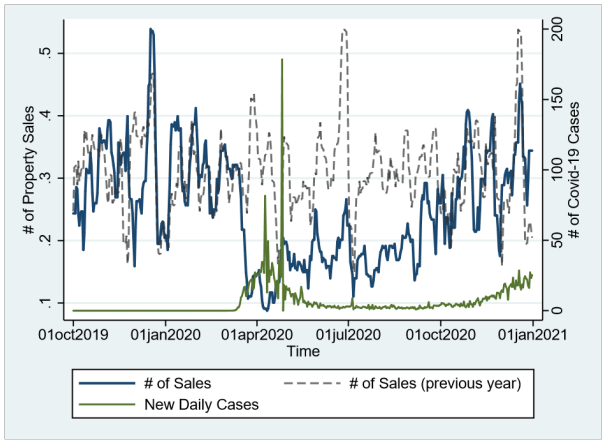
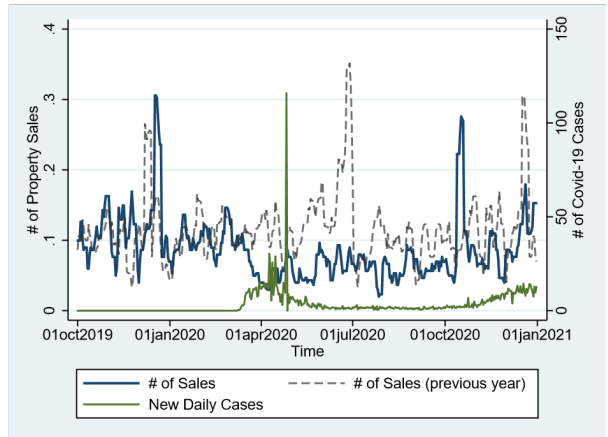

(a) Avg. MODZCTA Sale Price $< 25th$ perc.  (b) Avg. MODZCTA Sale Price $25th - 50th$ perc.

(c) Avg. MODZCTA Sale Price $50th - 75th$ perc.  (d) Avg. MODZCTA Sale Price $> 75th$ perc.

Figure 2: NYC Daily Case Counts and Avg. # of House Sales by Avg. MODZCTA Sale Prices

where $P_{it}$ represents the vector of sale prices of properties $i = 1, 2, .., N$ at time $t$, $\beta_0$ is an intercept term and $H_i$ represents a matrix of time-invariant property characteristics, such as the year a home was built, its square footage, and number of floors, among others.[6] $N_{bt}$ represents neighborhood attributes, e.g., the share of investor-owned properties at time $t$ in the MODZCTA $b$ where property $i$ is located. The main variables of interest intended to capture the pandemic-induced *contagion* and *income* effects on home sale prices are, respectively, given by $CC_{bt}$ and $E_{bt}$. While $E_{bt}$ represents economic factors, e.g., unemployment at time $t$ in the MODZCTA $b$ where property $i$ is located, $CC_{bt}$ represents the cumulative local Coronavirus infections per resident measuring the severity of a property's exposure to local cases in neighborhood $b$ at the time of sale. Neighborhood characteristics that are constant over time, such as distance to the nearest airport, are captured by a neighborhood fixed effects $\alpha_b$. Finally, $\alpha_t$ represents a matrix of time-of-sale fixed effects to capture citywide differences over time, such as overall general appreciation in property values, and $\varepsilon_{it}$ is an error term that is i.i.d. with mean zero and constant variance, along with zero co-variance across observations $i$, where $i = 1, 2, .., N$, and $N$ is the number of houses in the sample. One of the coefficients of interest is given by $\beta_1$ and measures the average change in house sale prices in response to a rise in cumulative local Coronavirus infections conditional on the other control variables, all else equal. The other coefficient of primary interest is given by $\varphi$ and measures the change in sale price per a one-unit change in neighborhood-level unemployment, ceteris paribus.

---

[6] While the PLUTO$^{TM}$ dataset contains a number property characteristics, it does not report all of the desirable home attributes, such the number of bedrooms and number of bathrooms.

The primary issue with this specification is that unobservable home and/or neighborhood characteristics of sold properties may correlate with the intensity of the local COVID-19 outbreak and bias our coefficient estimate of interest. If, for example, more affluent homeowners in any given neighborhood flee the City to avoid the pandemic, the neighborhood-specific composition of sold homes changes with the pandemic and potentially undermines our COVID-19 price effect estimate. Taking advantage of the repeat-sales sample, we can control for time-invariant observable and unobservable home and neighborhood characteristics by taking the first difference of Equation (2) between the initial and repeat sales. The basic repeat-sales approach was originally developed and popularized by Bailey *et al.* (1963) and popularized more broadly by Case and Shiller (1987). Since not all home and neighborhood characteristics are observable, this approach reduces the risk of omitted variable bias and inconsistent estimates (Chau and Chin, 2003) and can help to identify the COVID-19 price effect at the property rather than neighborhood level. As a result, the repeat-sales sample estimate is no longer sensitive to the composition of homes sold within a given neighborhood and much less prone to be biased.[7]

Specifically, suppose the second sale of a property occurs in time period $t + \tau$. This is a slightly varied version of equation (1), which is written as follows:

---

[7] That said, the repeat-sales estimator is still subject to a few concerns, such as changes in property attributes over time. We conduct a host of robustness analysis to scrutinize our repeat-sales identification strategy. The results of these sensitivity tests are in Section 5.3 (see Table 5) and produce consistent estimates that lend support to our identification strategy.

$$P_{i,t+\tau} = \beta_0 + \beta_1\, CC_{b,t+\tau} + \gamma H_i + \varphi E_{b,t+\tau} + \omega N_{b,t+\tau} + \alpha_b + \alpha_{t+\tau} + \varepsilon_{i,t+\tau}. \quad (3)$$

Taking the first difference (Δ) of Equation (2) and (3) for two separate sale dates for property *i*, which is sold at both time $t + \tau$ and time $t$ (where $t$ represents the initial sale and $t + \tau$ represents the subsequent sale), yields:

$$\Delta P_{i(t,t+\tau)} = \beta_1 \Delta\, CC_{b(t,t+\tau)} + \varphi \Delta E_{b(t,t+\tau)} + \omega \Delta N_{b(t,t+\tau)} + \Delta \alpha_{(t,t+\tau)} + \Delta \varepsilon_{(it,t+\tau)}. \quad (4)$$

The time-invariant characteristics of house *i*, $H_i$, and neighborhood *b*, $\alpha_b$, drop out of (4) when taking the first difference of Equations (2) and (3). Furthermore, time-of-home-sale and repeat-sale fixed effects ($\alpha_t$ and $\alpha_{t+\tau}$) control not only for the specific timing of each sale of a given house, but also the difference in the time elapsed between the two sales. That is, we would expect there to be a pandemic-unrelated difference in home value appreciation between a house sold 10 years ago and one sold 2 years ago. Similarly, we expect there to be a difference in the change in property prices for a home that was sold in 2010 and 2018 versus a home that was sold in 2012 and 2020. Our fixed effects capture the timing of both sales as well as the duration of time that has passed between repeat sales and therefore control for both of these potentially confounding factors.

The coefficients of interest are still given by $\beta_1$ and $\varphi$, but the interpretations change slightly. Given the specification in first differences, $\beta_1$ and $\varphi$ respectively quantify the average discount or premium on the change in sale prices of repeat-sales homes corresponding to a one-unit rise in novel Coronavirus infection and unemployment rates near a given property. A priori, we expect the severity of the pandemic to be a significant disamenity – both in terms of potential contagion and an adverse income shock – to home buyers that leads to a decline in average sale prices (or smaller home value appreciation between sales) in neighborhoods with rising Coronavirus infections and greater unemployment ($\beta_1 < 0$ and $\varphi < 0$).

Finally, we acknowledge the potential market segmentation across unobservable determinants (Adair et al., 1996; Fletcher et al., 2000) and expand our empirical analyses by investigating whether the COVID-19 outbreak has heterogeneous price effects across the top and bottom quartiles of potential home values and specific neighborhood attributes. We also present a set of robustness checks for the hedonic repeat sales analyses in Section 5.

# 4 Data

To estimate the correlations between COVID-19 and NYC property sale prices, we construct a novel dataset that combines the relevant information from multiple sources. The housing data, for example, was created through two publicly available resources from NYC government agencies. The first is the Primary Land Use Tax Lot Output (PLUTO$^{TM}$) database maintained by the NYC's Department of City Planning.[8] The PLUTO file contains detailed information on property characteristics by individualized tax lot. The second dataset entails information on the timing and price of every NYC property sale and is publicly available through the NYC's Department of Finance.[9] The two files share a unique identifier field in borough-block-lot (BBL) and may be merged once compiled.[10] Notably, the housing data includes both residential and commercial properties, the majority of which are reported sales of one- or two-family homes. It is these one- and two-family homes that are the focus of our analyses. The combined housing records span from January, 2003 through December, 2020. Other recent research on NYC property values has relied on the PLUTO dataset, including among others, Cohen *et al.* (2021).

In an attempt to isolate market arms-length transactions and minimize potential data errors, we employ a number of standard adjustments to these housing data. First, we remove all transactions with a sale price below $10,000. Next, we exclude all records with zero buildings, zero floors, zero total units, zero lot depth or zero square footage. In the event a property's designated land use changed during our sample or a sale was classified as a residential transaction, but the property has no residential units, it was removed. All records with a year built or year altered before 1800 were removed. We further exclude all observations pertaining to multiple sales of the same property on the same day.[11]

Finally, given that some single-family and two-family residential properties in NYC are rentals, we develop estimates of non-investor-owned properties. This is accomplished by following a methodology similar to that of Cohen and Harding (2020), where flags in the owner name field are used to filter out likely investor-owned residential properties. For instance, if the owner name for a property contains one of the terms "LLC", "CO", "INC", "DBA", among others, the owner is flagged as an investor. These flagged observations are excluded from the estimation samples. Moreover, we utilize this investor variable to determine the annual share of investor-owned properties among the stock of all properties within each MODZCTA. One potential concern is that the eviction moratorium, placing limitations on landlords evicting non-paying tenants, may disproportionately affect home values in neighborhoods with high rates of rental shares. If rental shares correlate with local case counts, estimates may be biased. Controlling for the share of investor-owned properties alleviates this concern.

---

[8] https://www1.nyc.gov/site/planning/data-maps/open-data/dwn-pluto-mappluto.page
[9] https://www1.nyc.gov/site/finance/taxes/property-rolling-sales-data.page
[10] Variations of these data have been used in the past (see, for example, Barr and Cohen (2014)).
[11] For some transactions these duplicate records may be plausible, perhaps indicating the simultaneous sale of multiple apartments or condominiums. For our sample of one- or two-family homes, however, these duplicate records cannot be easily reconciled. We error on the side of caution and exclude these observations, but note that our results are not sensitive to this (or any of the other) sample restrictions.

Table 1: NYC Borough One- and Two-Family Housing Market and Pandemic Summary Statistics

|  | (1) | (2) | (3) | (4) | (5) | (6) | (7) | (8) | (9) | (10) |
|---|---|---|---|---|---|---|---|---|---|---|
|  | Before Outbreak (03/01-12/31/2019) | | | After Pandemic Outbreak (03/01-12/31/2020) | | | | | | |
| Borough | Total # of Sales | Avg. Sale Price ($ '000) | Unemp. Rate (%) | Total # of Sales | Avg. Sale Price ($ '000) | Unemp. Rate (%) | Total # of Cases | Total # of Deaths | Max. Case Rate | Max. Death Rate |
| Brooklyn | 2,956 | 1,035.5 | 3.9 | 2,244 | 1,074.4 | 13.6 | 114,907 | 6,094 | 4,488.7 | 238.1 |
| Bronx | 1,345 | 567.0 | 5.1 | 1,049 | 598.0 | 17.3 | 80,347 | 4,175 | 5,654.2 | 293.8 |
| Manhattan | 65 | 3,228.3 | 4.0 | 53 | 4,521.8 | 12.9 | 55,104 | 2,673 | 3,419.6 | 165.9 |
| Queens | 5,637 | 738.0 | 3.3 | 4,095 | 761.1 | 13.4 | 116,779 | 6,323 | 5,148.2 | 278.8 |
| Staten Island | 3,049 | 595.7 | 3.7 | 2,506 | 617.3 | 11.3 | 32,589 | 1,102 | 6,844.4 | 231.4 |
| **NYC Average** | - | 766.9 | 3.7 | - | 798.4 | 14.7 | - | - | - | - |
| **NYC Total** | 13,052 | - | - | 9,947 | - | - | 399,726 | 20,367 | - | - |

Notes: Data on the number of one- and two-family property transactions and average sale prices are published by the NYC Department of Finance. Information on NYC borough unemployment rates are published by the Federal Reserve. Statistics on COVID-19 cases and deaths are publicly available from NYC Department of Health and reported on a daily frequency starting February 29, 2020. Based on our sample, we define the post outbreak period from March 1, 2020 until the end of our sample December 31, 2020. For the purposes of comparison, we calculate preoutbreak statistics on the number of sales and average property prices based on limited sample restricted to March 1 to December 31, 2019.

In aggregate, this leaves 306,508 total NYC property transactions of 269,478 one- or two-family homes between January, 2003 and July, 2020. Of these, we observe 9,947 sales during the COVID-19 period from March to December 2020 (see Table 1). In terms of repeat-sales homes, the data show a total of 73,568 non-investor transactions for 37,196 unique one- or two-family NYC properties since January, 2003. Of these, we observe 921 sales during the first year of the pandemic from March to December, 2020.

To provide initial evidence of the likely pandemic relationship with the NYC real estate market, we compare year-over-year changes in average sale prices and the number of transactions before and after the initial COVID-19 outbreak between March 1 and December 31, 2019 relative to 2020 (see columns (1) through (4) of Table 1). Taking the difference between columns (1) and (3) of Table 1, for example, shows that the total number of transactions of NYC one- and two-family homes decreases by 3,105 year-over-year and that this 24% reduction in the frequency of sales after the outbreak is not driven by a single borough, but the combined result of a notable housing market downturn in each of the five districts. Across NYC boroughs, the number of sales declined anywhere from 12 (Manhattan) to 1,542 (Queens) or 18% (Staten Island and Manhattan) to 27% (Queens) in relative terms.

Interestingly, the change in average sale prices does not reflect this pattern. In aggregate, the average property sale price from March 1 to December 31 rose by 4% from $766,900 to nearly $800,000. The increase in sale prices is consistent across all five boroughs (compare columns (2) and (4) of Table 1) and ranges from 3.1% in Queens to 40% in Manhattan.

To investigate whether these adjustments in the sale prices correlate with the local rise of the COVID-19 infections and the contemporaneous rise in unemployment, we combine these housing data with spatially disaggregated information on the NYC outbreak of the novel Coronavirus as well as unemployment rates. With respect to local unemployment rates we rely on two sources. The first is unemployment data generated by The DEEP-MAPS Project[12]. The DEEP-MAPS model combines information from the Current Population Survey (CPS) and the Local Area Unemployment Statistics (LAUS) data sources to produce a complete national database of inferred labor force participation at the census-tract level. The DEEP-MAPS data is available at a monthly frequency for the calendar year 2020.

The second source provides monthly pre-pandemic unemployment statistics through February 2020 at the borough level. These unemployment rates are published by the U.S. Bureau of Labor Statistics retrieved from the Federal Reserve Bank of St. Louis (FRED). We use the overlap in census tract and borough-level data from January and February, 2020 to backcast[13] local unemployment rates based on the available borough data. Our results are robust to this extrapolation.

Table 1 shows that the average borough-level pre-pandemic unemployment rate from March through December 2019 ranged from 3.3% in Queens to 5.1% in the Bronx and averaged 3.7% for all of NYC (see column (3)). One year later, NYC unemployment rates rose more than threefold during the first and second wave of the pandemic. Column (6) of Table 1 shows average unemployment rates ranging from 11.3% in Staten Island to 17.3% in the Bronx; with an NYC average of 14.7%.

---

[12] The Demographic Estimates of Employment Participation, with Multistage Adjustment and Poststratification Synthetics (DEEP-MAPS) model of the labor force projects labor force statistics to small subgroups of the population. For additional information, please see (Ghitza and Steitz, 2020). Source: https://www.deepmaps.io/data

[13] We backcast using the DEEP-MAPS and FRED borough level to get estimates at the MODZCTA level for years 2003-2019.

With respect to pandemic-related data, the NYC Department of Health publishes the number of positive tests for COVID-19, the number hospitalized COVID-19 patients, and the number of COVID-19 related deaths as well as their respective rates per 100,000 residents on a daily basis. The earliest reporting dates back to February 29, 2020 and provides these statistics at the NYC borough level differentiating, for example, between Brooklyn and Manhattan case counts and rates. Starting on April 3, 2020, the NYC Department of Health began to publish pandemic statistics on the total number of tests and positive test results at the more disaggregated modified zip code (MODZCTA) level. By May 18, 2020, the agency augmented these data with information on death counts and death rates per 100,000 MODZCTA residents. We use the overlap in MODZCTA- and borough-level data from April 3 through December 31, 2020 to backcast case counts at the MODZCTA level through March 1, 2020 based on the available borough counts.[14] Our results are robust to this extrapolation.

In total, there are 176 MODZCTA's scattered throughout the five NYC boroughs including the Bronx, Brooklyn, Manhattan, Queens, and Staten Island. We build a MODZCTA-to-BBL concordance and match each of the sold properties to its corresponding MODZCTA and borough. In addition to the aforementioned real estate statistics, Table 1 also summarizes pandemic information at the borough level (see columns (7) through (10)). Column (7) indicates that nearly 400,000 people residing in NYC contracted the Coronavirus by December 31, 2020, most of whom live in the Bronx, Brooklyn, and Queens. Of these individuals who tested positive in NYC, over 20,000 died. Brooklyn and Queens lead these statistics with over 6,000 infected individuals who perished (see column (8)). These aggregate pandemic statistics are notably lower for Manhattan and Staten Island than those observed in Brooklyn and Queens. Staten Island, for example, appears much less severely affected, often reporting case and death counts four to five times lower than those observed in the other boroughs. This distinction, however, becomes less obvious when taking borough-level population into account. Considering case and death rates per 100,000 borough residents, all boroughs experienced the COVID-19 outbreak with similar severity by December 31, 2020 (see columns (9) and (10) of Table 1).

---

[14] Specifically, we calculate the MODZCTA share in total borough cases when both case counts are available and use these shares to predict MODZCTA case counts from March 1 through April 3, 2020, when only borough-level data are available.

While Table 1 provides important first insights into the potential pandemic effects on housing markets across the five boroughs, the aggregated statistics through December 2020 mask a few interesting facts. Table 2 offers a more detailed picture of the likely pandemic-related changes in the NYC real estate market. Specifically, we present pre- and post-pandemic sample averages in sale prices, sale volumes, unemployment rates, property attributes, and neighborhood characteristics during the first wave of COVID-19 infections through July 31, 2020 (Panel A of Table 2) and those observed during the onset of the second wave of infections through December 31, 2020 (Panel B of Table 2). We test for differences in these means and differentiate the characteristics for the sub-sample of repeat-sales properties (columns (6) through (10)).

MODZCTA neighborhood characteristics are based on the 2019 American Community Survey (ACS) published by the U.S. Census Bureau. The reported pre- and post-pandemic averages represent the average characteristics associated with the observed property sales and vary with the composition of sales through time. Therefore, the reduction in sample average neighborhood density per unit during the second wave of infections (see columns (5) and (10) of Panel B of Table 2), for example, should not be interpreted as a change in population density in NYC between 2019 and 2020, but instead should be interpreted as a change in the location of a typical property sale towards less (pre-pandemic) densely populated neighborhoods in NYC.

The statistics reported in Table 2 reveal several interesting patterns. First, the pre- to post-pandemic dynamics of the NYC real estate market are distinctly different during the first wave of the novel coronavirus outbreak relative to its second wave. Excluding repeat sales properties, Table 2, for example, shows that the change in average sale prices is negligible during the first wave of the pandemic, but is positive and statistically significant during the second wave of infections, albeit the average characteristics of these homes do not seem to change relative to pre-pandemic levels. Table 2 also demonstrates that the average rise in local unemployment rates (relative to pre-pandemic levels) is less pronounced during the second wave of infections. This partial recovery in unemployment rates correlates with a positive change in the volume of real estate transactions. While the market experiences a significant drop in the volume of sales during the first wave of the COVID-19 outbreak (irrespective of neighborhood characteristics), the number of real estate transactions per MODZCTA notably recovers during the latter half of 2020. Moreover, this recovery in property sales appears correlated with certain neighborhood attributes. Table 2 shows that during the second wave of the pandemic a disproportionate number of sales occurs in higher-income neighborhoods with more White and college-educated residents. Furthermore, recovery seems to be driven by sales in less densely populated neighborhoods with fewer rental units.

The second interesting pattern revealed by Table 2 pertains to differences across one-time sales (columns (1) through (5)) and repeat-sales properties (columns (6) through (10)). Although there are significantly fewer transactions of repeat-sales properties, the changes in property characteristics for repeat-sales homes tend to be very similar to those observed for homes that are sold only once over the sample duration. That is, the transactions of both one-time sales and repeat-sales properties experience a significant reduction in the volume of sales and face significantly higher local unemployment rates during the first wave of infections. Other property and location characteristics appear to remain stable during the first wave of the pandemic, irrespective of whether it is the first or a repeat transaction of a given property. The same tends to be true during the second wave of infections, where we note significant changes in location characteristics and show that these differences are similar for one-time and repeat-sales homes.

An exception to these similarities between one-time and repeat-sales homes is the average change in sale prices during the first wave of infections. While properties with a single reported sale show no statistically significant change in sale prices between the first half of 2019 and the first half of 2020, the average sale price of repeat-sales properties is shown to rise significantly in early 2020 relative to pre-pandemic levels.

To take a closer look at the spatial patterns of the pandemic, we map MODZCTA COVID-19 cumulative case rates as of July 31, 2020, towards the end of the first wave of infections, as well as December 12, 2020, in the midst of the second wave of infections. Figures 3a through 3b reiterate our initial observations. First, the severity of the outbreak is greater in the Bronx, Queens, and Staten Island where the majority of MODZCTAs report cumulative case rates above the $50^{th}$ percentile of the NYC COVID-19 case rate distribution. In contrast, many of Manhattan's MODZCTAs experience cumulative cases rates at or below the $25^{th}$ percentile irrespective of first or second wave. On the other hand, Figures 3a through 3b also show that the virus had spread far throughout NYC with a minimum MODZCTA case rate of 663 as of July 31, 2020 and a minimum of 1,637 cases per 100,000 residents as of December 31, 2020.

Second, Figures 3c and 3d show that the spatial distribution of the year-over-year change in the number of one- or two-family home sales tends to negatively correlate with the number of COVID-19 cases mapped in Figures 3a and 3b. This is especially true during the first wave of infections. From March 1, 2020 through July 31, 2020, Staten Island, Brooklyn, and Queens experience some of the largest reductions in the number of transactions. MODZCTAs in Manhattan report the smallest year-over-year reductions, and even some increases, in the number of property sales, though it is notable that Manhattan also has the smallest number of one or two-family homes. This inverse correlation between the volume of sales and cumulative case rates is still apparent (i.e., see Manhattan, and parts of the Bronx and Queens), but not as strong during the second wave of infections. Particularly, Staten Island and parts of Brooklyn defy this pattern demonstrating larger sales volumes and higher cumulative case rates.

Third, year-over-year changes in the average MODZCTA sale prices, depicted by Figures 3e and 3f, are much more volatile and more dispersed across NYC MODZCTA's than Coronavirus case rates and are therefore only mildly and negatively correlated. Most MODZCTAs in Manhattan report zero sales during the COVID-19 period in 2020 and therefore we cannot determine the change in average sale prices. For those Manhattan MODZCTAs that report sales during the relevant periods in 2019 and 2020, the average sale price of one- or two-family homes either falls substantially, by as much as $12.4 million, or increases dramatically by up to $29 million. Across the remaining boroughs, around two-thirds of MODZCTAs experience a modest increase in average sale prices (see Figures 3e and 3f).

## 5 Results

Our empirical analysis estimates the COVID-19 case effect on the sale price of NYC one- and two-family homes. We differentiate the pandemic effects across the full sample of one- and two-family homes as well as the restricted repeat-sales sample. Moreover, the analyses provide estimates of the pandemic effects during the first wave of infections (samples restricted to the months of March through July of any given year to control for seasonal effects) and the second wave of infections (samples restricted to the months of August through December of any given year to control for seasonal effects). The results point to statistically significant *contagion* and *income* effects on the NYC real estate market. The estimates notably vary across the first and second waves of infections and differ for repeat sales.

### 5.1 Baseline Price Analysis

We begin our price analysis with the standard hedonic regression on the full sample of one- and two-family home sales (non-investor-owned). Table 3 shows we differentiate the effects between the first wave of infections, including sales data from March through July from 2003 through 2020, and the second wave of infections, including sales data from August through December from 2003 through 2020. We restrict the samples to these five-months windows per year to control for seasonality in the real estate market. First wave estimates are shown in columns (1) and (2), whereas second wave estimates are given in columns (5) and (6). Each regression controls for the local neighborhood share of investor-owned single- and two-family residential properties as well as available property characteristics, such as the age of the home, its building and lot square footage, number of floors, and whether the property has a basement and has been altered since 2000.[15] Moreover, we include MODZCTA and month-of-sale fixed effects to control for other unobservable time-invariant neighborhood attributes and common trends.

The coefficient estimates of interest suggest that a rise in the rate of cumulative MODZCTA Coronavirus infections per 100,000 residents is associated with a significant reduction NYC home sale prices. This effect is found to be notably larger during the second wave of the pandemic compared to the first (compare columns (1) and (2) vs. (5) and (6) of Table 3) and robust to the inclusion of the neighborhood unemployment rate. As we expected, a rise in unemployment significantly lowers property sale prices and the estimated effect is relatively stable across both estimation samples (see columns (2) and (6)).

Coefficient estimates for the control variables tend carry the expected sign (Chau and Chin, 2003) and some are found to be statistically significant at the 1% to 10% levels (see columns (1), (2), (5), and (6) of Table 3). Among the statistically significant estimates, we find that property modifications are associated with increases in property values. For every alteration since the year 2000, the average sale price of one- and two-family homes rises by approximately $135,000. Similarly, larger homes located on larger lots and those with basements command a price premium adding, for example, over $210,000 in value per 1,000 additional square feet of living space.[16] The share of investor-owned properties within a given home's neighborhood does not seem to be strongly correlated with its sale price.

---

[15] According to the meta data of the PLUTO™ dataset, the NYC Department of Finance defines alterations as modifications to the structure that, according to the assessor, change the value of the real property. The year of alteration is based on the associated building permits.

[16] The PLUTO™ dataset includes a number of potential control variables that describe the characteristics of a sold property. Among the possibilities, we integrate structural home attributes that are consistently available for our sample of sold homes and have been shown to be relevant determinants of house sale prices in the previous literature (see, for example, Chau and Chin (2003)). Our results are quantitatively and qualitatively robust to the inclusion of alternative characteristics, such as indicator variables for NYC neighborhood or health area, among others, and are available upon request.

Table 2: Changes in Pre- and Post-Pandemic NYC Real Estate Market Characteristics

| | (1) | (2) | (3) | (4) | (5) | (6) | (7) | (8) | (9) | (10) |
|---|---|---|---|---|---|---|---|---|---|---|
| | \multicolumn{5}{c}{Sales of once-sold Properties} | \multicolumn{5}{c}{Sales of repeatedly-sold Properties} | | | | | |
| Characteristic | Pre-Mean | Pre-S.D. | Post-Mean | Post-S.D. | Δ in Means | Pre-Mean | Pre-S.D. | Post-Mean | Post-S.D. | Δ in Means |
| **Panel A: First Wave of Infections (03/01/2019-07/31/2019 vs. 03/03/2020-07/31/2020)** | | | | | | | | | | |
| Sale Price ($ '000) | 784.74 | (600.4) | 793.64 | (579.2) | 8.90 | 660.50 | (470.3) | 763.98 | (384.9) | 103.48*** |
| # of Sales | 81.20 | (60.3) | 49.43 | (34.7) | -31.78*** | 10.24 | (6.5) | 5.09 | (3.2) | -5.15*** |
| Unemployment Rate (%) | 4.04 | (2.8) | 14.07 | (8.1) | 10.03*** | 5.13 | (3.4) | 15.48 | (8.1) | 10.35*** |
| Renovated Home (yes==1) | 0.04 | (0.2) | 0.05 | (0.2) | 0.01 | 0.03 | (0.2) | 0.04 | (0.2) | 0.01 |
| Building Area ('00 sqft) | 24.34 | (164.6) | 28.95 | (222.0) | 4.61 | 16.57 | (6.2) | 17.52 | (7.4) | 0.95 |
| Lot Area ('000 sqft) | 7.25 | (109.4) | 10.93 | (162.4) | 3.68 | 2.88 | (1.5) | 2.90 | (1.9) | 0.02 |
| # of Floors | 2.03 | (0.5) | 2.03 | (0.5) | -0.00 | 2.04 | (0.5) | 2.09 | (0.5) | 0.05 |
| Basement (yes==1) | 1.00 | (0.0) | 1.00 | (0.0) | 0.00 | 1.00 | (0.0) | 1.00 | (0.0) | 0.00 |
| Population ('000) | 53.94 | (25.2) | 54.89 | (25.3) | 0.95 | 54.27 | (24.9) | 57.51 | (24.0) | 3.24 |
| Share of White Residents (%) | 39.46 | (17.1) | 38.05 | (16.9) | -1.41*** | 38.56 | (18.3) | 38.96 | (18.0) | 0.40 |
| Share of Rental Units (%) | 45.95 | (19.9) | 46.45 | (20.0) | 0.50 | 47.16 | (20.5) | 48.07 | (21.5) | 0.90 |
| Density per Unit | 2.79 | (0.4) | 2.78 | (0.4) | -0.02 | 2.88 | (0.4) | 2.86 | (0.4) | -0.03 |
| English Proficiency | 11.71 | (10.2) | 11.73 | (10.2) | 0.02 | 9.73 | (8.9) | 10.90 | (9.9) | 1.17 |
| Median Income ($ '000) | 73.67 | (18.6) | 72.98 | (18.2) | -0.69 | 70.26 | (19.0) | 69.64 | (20.5) | -0.61 |
| Share College Educated (%) | 56.33 | (8.9) | 56.30 | (8.8) | -0.03 | 53.91 | (8.9) | 53.85 | (9.9) | -0.06 |
| Observations | 5,670 | | 3,471 | | 9,141 | 542 | | 277 | | 819 |
| **Panel B: Second Wave of Infections (08/01/2019-12/31/2019 vs. 08/01/2020-12/31/2020)** | | | | | | | | | | |
| Sale Price ($ '000) | 769.41 | (517.7) | 806.35 | (754.2) | 36.94** | 647.07 | (381.2) | 740.55 | (483.4) | 93.48** |
| # of Sales | 89.37 | (63.4) | 90.64 | (74.8) | 1.27 | 9.26 | (6.4) | 8.79 | (9.2) | -0.47 |
| Unemployment Rate (%) | 3.59 | (2.5) | 12.09 | (4.0) | 8.50*** | 4.41 | (2.9) | 12.59 | (3.7) | 8.18*** |
| Renovated Home (yes==1) | 0.04 | (0.2) | 0.04 | (0.2) | 0.00 | 0.02 | (0.1) | 0.03 | (0.2) | 0.01 |
| Building Area ('00 sqft) | 24.42 | (170.0) | 28.29 | (211.5) | 3.87 | 32.48 | (258.6) | 29.69 | (243.7) | -2.80 |
| Lot Area ('000 sqft) | 7.76 | (120.8) | 9.87 | (143.7) | 2.11 | 12.05 | (166.3) | 13.50 | (190.8) | 1.45 |
| # of Floors | 2.03 | (0.5) | 2.02 | (0.5) | -0.01 | 2.00 | (0.5) | 2.02 | (0.5) | 0.02 |
| Basement (yes==1) | 1.00 | (0.0) | 1.00 | (0.0) | 0.00 | 1.00 | (0.0) | 1.00 | (0.0) | 0.00 |
| Population ('000) | 54.49 | (25.6) | 52.82 | (25.2) | -1.66*** | 53.94 | (24.5) | 54.13 | (24.1) | 0.19 |
| Share of White Residents (%) | 39.38 | (17.0) | 40.96 | (17.8) | 1.58*** | 40.09 | (18.6) | 43.97 | (19.2) | 3.89** |
| Share of Rental Units(%) | 46.06 | (19.8) | 44.28 | (20.1) | -1.78*** | 46.22 | (20.1) | 42.67 | (20.8) | -3.55* |
| Density per Unit | 2.79 | (0.4) | 2.76 | (0.4) | -0.03*** | 2.86 | (0.4) | 2.79 | (0.4) | -0.06* |
| English Proficiency | 11.88 | (10.4) | 11.64 | (10.4) | -0.24 | 9.88 | (9.2) | 9.51 | (8.4) | -0.37 |
| Median Income ($ '000) | 73.64 | (18.4) | 75.43 | (18.8) | 1.79*** | 71.62 | (19.4) | 75.82 | (19.2) | 4.20** |
| Share College Educated (%) | 56.39 | (8.6) | 57.00 | (8.8) | 0.62*** | 54.67 | (8.8) | 56.56 | (8.7) | 1.89** |
| Observations | 6,345 | | 5,845 | | 12,190 | 495 | | 354 | | 849 |

Notes: Data on the number of one- and two-family property transactions and sale prices are published by the NYC Department of Finance. Information on NYC MODZCTA unemployment rates are published by DEEP-MAPS. Statistics on housing characteristics are publicly available through the PLUTO$^{TM}$ database maintained by the NYC Department of City Planning. Neighborhood characteristics represent the sample averages for the respective samples of observed sales and are based on 2019 data available through American Community Survey published by the U.S. Census. Based on our sample, we define distinct post outbreak periods including the 1) the 1$^{st}$ wave of infections from March 1, 2020 through July 31, 2020; and 2) 2$^{nd}$ wave of infections from August 1, 2020 through December 31, 2020. For the purposes of comparison, we calculate pre-outbreak statistics based on a limited sample restricted to 1) March 1 to July 31, 2019; and 2) August 1 to December 31, 2019.

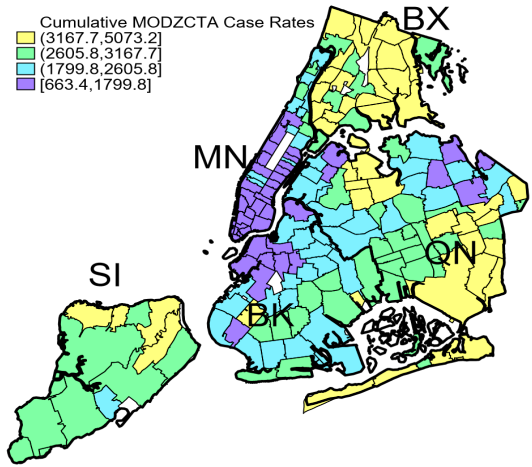
(a) Cumulative MODZCTA Case Rates (July 31, 2020)

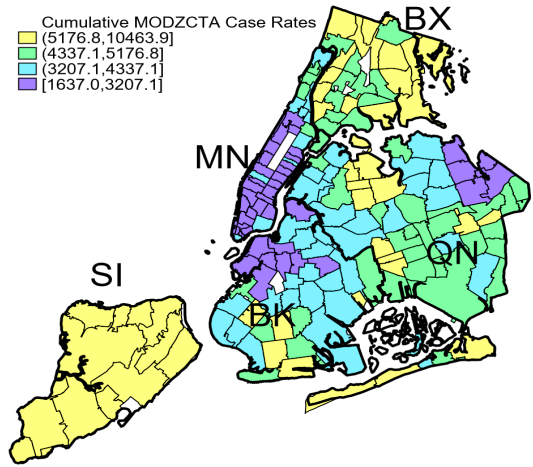
(b) Cumulative MODZCTA Case Rates (Dec. 31, 2020)

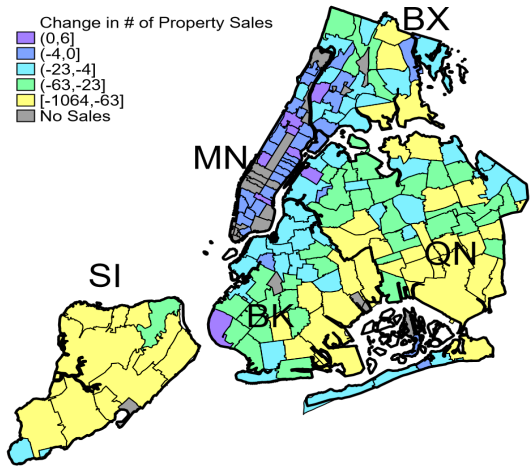
(c) YoY Δ in # of Sales (03/01-07/31, 2019 to 2020)

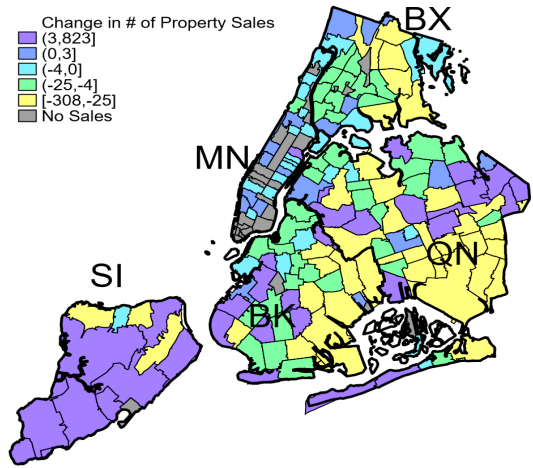
(d) YoY Δ in # of Sales (08/01-12/31, 2019 to 2020)

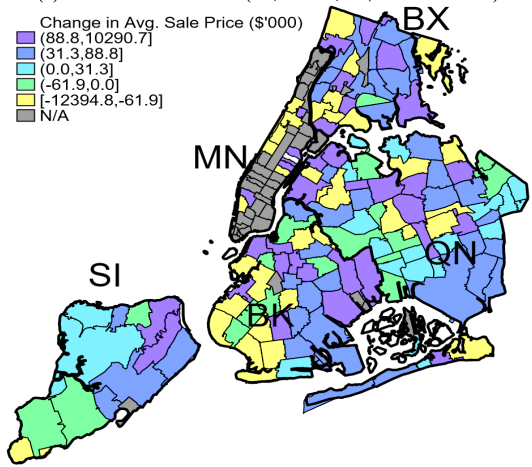
(e) YoY Δ in Avg. Prices (03/01-07/31, 2019 to 2020)

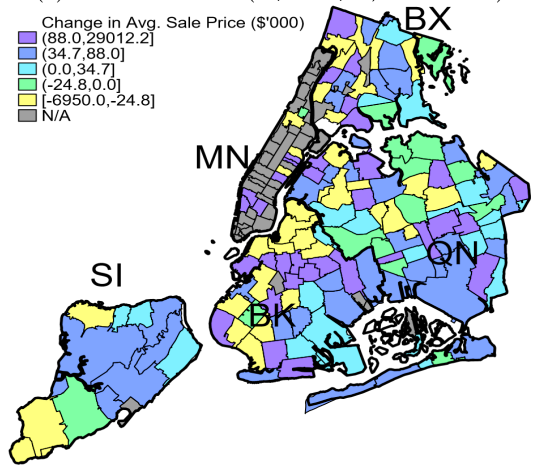
(f) YoY Δ in Avg. Prices (08/01-12/31, 2019 to 2020)

Figure 3: Cumulative Case Rates and MODZCTA Changes in the # of Property Sales and Avg. Sale Prices by Wave of Infection

While these initial estimates are insightful, it is plausible that COVID-19 infections and higher unemployment rates are simply concentrated in lower-valued neighborhoods. The repeat-sales sample overcomes this issue by differencing out unobservable property-specific characteristics, such as whether a home is a relatively high- or low-valued property within a given neighborhood, which may be uncorrelated with the observable characteristics we previously included. The repeat-sales estimate of the COVID-19 effect is therefore identified through the variation in price and COVID-19 exposure at the property level, rather than changes in the average price across a given neighborhood, where the sample of sold homes may have been altered by the pandemic itself. Controlling for time-varying neighborhood and property attributes that cannot be differenced out, such as changes in unemployment, shares of investor-owned properties, and home modifications, we interpret the repeat-sales estimate as the relationship between COVID-19 and NYC residential property prices.

These pandemic effect estimates based on the repeat sales sample indicate similar patterns during the second wave of the pandemic, but are less pronounced during the first (compare columns (3) and (4) vs. (7) and (8) of Table 3). Controlling for the months of initial sale and repeat sale as well as changes in the share of investor-owned properties within the neighborhood and property modifications between sales, we find that a rise in the cumulative case rate of 1,000 novel Coronavirus infections per 100,000 MODZCTA residents during the first wave of the pandemic is associated with a reduction in the average property sale price of approximately $70,000 (see column (3) of Table 3). A similar rise in the rate of new infections during the first outbreak is found to have negligible effects (see column (3) of Table 3). Similarly, unemployment is associated with a reduction of sale prices during the first wave of infections, but the estimates are statistically indistinguishable from zero.

In contrast, during the second wave of the pandemic increases in the cumulative and new case rates lead to significant reductions in sale prices of repeat-sales one- or two-family homes. Column (7) of Table 3, for example, shows that for a cumulative case rate increase of 1,000 infections, property prices between sales fall by $63,000 on average. As expected, these sale prices also respond negatively to the significant increases in local unemployment rates observed during the second half of 2020. A 10-percentage point increase in the local unemployment rate, for example, is found to reduce the average change property values of repeatedly sold homes by $60,000 (see column (7) of Table 3).

Again, we find that changes in the share of investor-owned properties do not influence non-investor-owned home sale prices, whereas property modifications between sales have a statistically significant

Table 3: COVID-19 Case Effect on NYC One & Two Family Home Prices

| | (1) | (2) | (3) | (4) | (5) | (6) | (7) | (8) |
|---|---|---|---|---|---|---|---|---|
| | \multicolumn{4}{c}{First Wave (03/01-07/31, 2003-2020)} | | | | |
| | \multicolumn{4}{c}{} | \multicolumn{4}{c}{Second Wave (08/01-12/31, 2003-2020)} |
| | All Sales | | Repeat Sales | | All Sales | | Repeat Sales | |
| | Cum. Cases | Cum. Cases | Cum. Cases | New Cases | Cum. Cases | Cum. Cases | Cum. Cases | New Cases |
| Cases per 100,000 ('000) | -0.382** | -0.398** | -0.706* | 10.162 | -0.611** | -0.618*** | -0.633*** | -30.759*** |
| | (0.172) | (0.174) | (0.358) | (12.208) | (0.236) | (0.235) | (0.241) | (7.905) |
| Unemployment Rate (%) | | -0.064*** | -0.039 | -0.039 | | -0.072*** | -0.060** | -0.063*** |
| | | (0.012) | (0.024) | (0.024) | | (0.013) | (0.024) | (0.024) |
| Investor Owned (%) | 0.004 | 0.004 | -0.009 | -0.009 | 0.002 | 0.002 | -0.011 | -0.010 |
| | (0.009) | (0.009) | (0.013) | (0.013) | (0.010) | (0.010) | (0.014) | (0.014) |
| Altered (yes==1) | 1.356*** | 1.351*** | 2.432*** | 2.432*** | 1.997*** | 1.989*** | 3.056*** | 3.058*** |
| | (0.228) | (0.228) | (0.339) | (0.340) | (0.315) | (0.315) | (0.591) | (0.591) |
| Year Built | 0.002 | 0.002 | | | 0.003 | 0.003 | | |
| | (0.002) | (0.002) | | | (0.002) | (0.002) | | |
| Home Sqft ('000) | 2.174*** | 2.163*** | | | 2.174*** | 2.159*** | | |
| | (0.280) | (0.281) | | | (0.341) | (0.341) | | |
| Lot Sqft ('00,000) | 0.032** | 0.032** | | | 0.027* | 0.026* | | |
| | (0.014) | (0.013) | | | (0.015) | (0.015) | | |
| # of Floors | 0.051 | 0.060 | | | 0.045 | 0.067 | | |
| | (0.069) | (0.071) | | | (0.159) | (0.158) | | |
| Basement (yes==1) | 0.519*** | 0.502*** | | | 0.530*** | 0.506*** | | |
| | (0.109) | (0.112) | | | (0.133) | (0.132) | | |
| Constant | -1.475 | -1.372 | 0.790*** | 0.765*** | -3.643 | -3.317 | 0.753*** | 0.722*** |
| | (3.674) | (3.633) | (0.085) | (0.083) | (4.459) | (4.413) | (0.072) | (0.069) |
| $N$ | 122,630 | 121,980 | 15,540 | 15,540 | 133,072 | 132,393 | 15,733 | 15,733 |
| adj. $R^2$ | 0.606 | 0.608 | 0.146 | 0.146 | 0.551 | 0.552 | 0.131 | 0.131 |
| F | 55.2*** | 47.4*** | 14.2*** | 13.4*** | 35.1*** | 32.3*** | 10.4*** | 11.5*** |
| Month-of-sale FE | Y | Y | Y | Y | Y | Y | Y | Y |
| Month-of-previous-sale FE | N | N | Y | Y | N | N | Y | Y |
| MODZCTA FE | Y | Y | N | N | Y | Y | N | N |

Notes: Standard errors are reported in parenthesis. Across all specifications, we cluster standard errors at the MODZCTA level. The estimation sample underlying the results presented in columns (1) through (4) focuses on the pandemic-induced effects during first wave of infections and includes transactions of one or two-family homes sold in NYC between March and July for each year from 2003 through 2020, whereas the estimation sample underlying the results shown in columns (5) and (8) focus on the second wave and includes transactions of one or two-family homes sold in NYC between August and December for each year from 2003 through 2020. Estimation samples underlying the results presented in columns (1) and (2) as well as (5) and (6) include all sales of one or two-family home in NYC and each variable is given in levels, i.e. the dependent variable is given by the recorded sale price in $100,000. In contrast, estimation samples underlying the results presented in columns (3) and (4) as well as (7) and (8) are restricted to repeatedly-sold one or two-family properties. Accordingly, all variables are given in changes between sales, i.e. the dependent variable is given by the change in sale prices between sales in $100,000. Depending on the sample restriction and selection of control variables, these estimation samples include 149 to 160 MODZCTAs. The cumulative and new case rates are defined per 100,000 MODZCTA residents. Across all estimations investor-owned properties are excluded. The variable "Investor Owned" controls for the share of investor-owned properties within a given home's MODZCTA. Statistical significance at the conventional 10%, 5%, and 1% thresholds is given by *, **, ***, respectively.

positive effect on resale values. Other time-invariant property characteristics variables drop out when taking first differences for the repeat-sales estimations.

Combined, the estimates reported in columns (1) through (8) suggest that larger COVID-19 case numbers occur in neighborhoods of lower-valued properties (see columns (1) and (2) as well as columns (5) and (6)) and that a worsening of the COVID-19 outbreak – especially during the second wave of infections – within any given neighborhood, when controlling for unemployment changes, is correlated with a reduction in the average value of repeat-sales properties within that MODZCTA (columns (3), (7) and (8)). These findings echo some of the recent reports on COVID-19, which have shown that the outbreak is more concentrated among poorer (i.e., unemployed) individuals (Goldstein, 2020), who are more likely to own or occupy lower-valued homes.

## 5.2 Price Heterogeneity Analysis

A priori, it is unclear whether these pandemic price effects are similar across price levels or how they may interact with buyer attributes and preferences. More affluent home buyers, for example, may be more sensitive to the possibility of contagion, while wealthier neighborhoods may be less likely to be impacted by contagion during the pandemic (i.e., less densely populated and more opportunity to isolate and work remotely). In the absence of observing buyer characteristics, we rely on pre-pandemic neighborhood attributes to test whether the *contagion* and *income* effects on home values vary with the characteristics of the residents living near the location of the sale. The pre-pandemic attributes are based on the 2019 ACS and include population density per BBL, median neighborhood income, the share of rentals among all residential units, and the local population shares of residents using public transit, with limited English proficiency, having some college education, or who are foreign born.

Table 4 provides the coefficient estimates for these interaction terms during the first wave of infections (Panel A) and the second wave of the pandemic (Panel B). Column (1) of Table 4 provides the baseline average effect estimates. In columns (2) and (3), we present price effect estimates for homes valued in the bottom and top $25^{th}$ percentile of the observed local pre-pandemic price distribution, respectively. The results reveal an interesting pattern. While the *contagion* price discount is orders of magnitude larger for the least valued properties during the first wave of infections, these properties face insignificant COVID-19 case effects during the second wave (see column (2)).[17]

The interactions with pre-pandemic neighborhood attributes are shown in columns (4) through (10) of Table 4 and reveal a number of interesting patterns regarding the pandemic-induced price effects. A priori, one might expect the *contagion* effect on prices to intensify in more densely populated neighborhoods that are more reliant public transit. The estimates shown in columns (4) and (5) point to the contrary. For repeatedly-sold properties located in neighborhoods with above median density per BBL and above median reliance on public transit, the negative price effect of a rise in Coronavirus infections vanishes. In contrast, the *income* price effect, measured through changes in local unemployment rates, intensifies in these types of neighborhoods. One way to interpret these results is that home buyers of properties located in more densely populated neighborhoods are less risk averse to the potential contagion but are more sensitive to changes in income.

Columns (6) through (8) illustrate pandemic price effect patterns with respect to pre-pandemic neighborhood income, education, and home ownership. In general, the estimates suggest that the *contagion* price effect intensifies among home buyers of properties located in more affluent neighborhoods, whereas the *income* effect is less pronounced in these locations. Median income and rental shares have particularly strong effects (see columns (6) and (8)). In neighborhoods with typical household earnings above the NYC median income the price discount of a rise in infections more than doubles during the first wave to the pandemic, while the effect of higher unemployment rates approaches zero. With respect to rental shares, we observe that the *contagion* effect is driven by price changes of repeatedly sold properties located in neighborhoods with below median rental shares, whereas the income price effect is negligible in these neighborhoods.

Finally, we test whether the estimated price effect varies with factors that may influence information accessibility. To this end, we interact case and unemployment rates with a dummy variable that indicates whether a repeat-sale property is located in a neighborhood with above median pre-pandemic shares of households with limited English proficiency or foreign-born residents. Home buyers with limited access to information regarding the severity of the local outbreak, for example, may be less influenced by rising

---

[17] The price effect of additional infections appears to intensify for high-valued properties during the second wave of the pandemic.

Table 4: Heterogeneity of COVID-19 Case and Income Effects on Sales Prices of Repeatedly-sold Homes

| | (1) Baseline Results | (2) Price < 25th Perc. | (3) Price > 75th Perc. | (4) BBL Density | (5) Public Transit | (6) Median Income | (7) College Educated | (8) Rental Share | (9) Limited English | (10) Foreign Born |
|---|---|---|---|---|---|---|---|---|---|---|
| | | | | \multicolumn{7}{c}{Interaction w. Neighborhood Attributes} | | | |
| **Panel A: First Wave of Infections (03/01-07/31, 2003-2020)** | | | | | | | | | | |
| Cases per 100,000 ('000) | -0.706* | -6.459*** | -0.854* | -0.848** | -0.891** | -0.591* | -0.745** | -0.972** | -1.012*** | -0.998*** |
| | (0.358) | (1.364) | (0.469) | (0.396) | (0.346) | (0.339) | (0.355) | (0.407) | (0.380) | (0.365) |
| Cases X Neigh. Attributes | | | | 0.867*** | 0.739*** | -0.693*** | -0.348 | 0.999*** | 0.531** | 0.501** |
| | | | | (0.252) | (0.247) | (0.234) | (0.212) | (0.251) | (0.241) | (0.219) |
| Unemployment Rate (%) | -0.039 | -0.032 | 0.019 | -0.005 | -0.020 | -0.053** | -0.042* | 0.014 | -0.038 | -0.005 |
| | (0.024) | (0.027) | (0.064) | (0.025) | (0.024) | (0.026) | (0.024) | (0.026) | (0.024) | (0.032) |
| Unemp. X Neigh. Attributes | | | | -0.080*** | -0.064** | 0.044* | 0.008 | -0.086*** | -0.015 | -0.055* |
| | | | | (0.025) | (0.028) | (0.023) | (0.027) | (0.023) | (0.025) | (0.025) |
| N | 15,540 | 2,915 | 4,360 | 15,540 | 15,540 | 15,540 | 15,540 | 15,540 | 15,540 | 15,540 |
| adj. R² | 0.146 | 0.168 | 0.093 | 0.147 | 0.147 | 0.147 | 0.146 | 0.148 | 0.146 | 0.146 |
| F | 14.2*** | 7.8*** | 16.3*** | 10.8*** | 11.3*** | 10.7*** | 10.6*** | 11.7*** | 10.3*** | 10.9*** |
| **Panel B: Second Wave of Infections (08/01-12/31, 2003-2020)** | | | | | | | | | | |
| Cases per 100,000 ('000) | -0.633*** | 0.200 | -1.242*** | -0.694*** | -0.806*** | -0.549** | -0.639** | -0.701*** | -0.712*** | -0.677*** |
| | (0.241) | (0.724) | (0.353) | (0.243) | (0.236) | (0.242) | (0.246) | (0.243) | (0.232) | (0.231) |
| Case X Neigh. Attributes | | | | 0.550*** | 0.636*** | -0.276*** | -0.095 | 0.482*** | 0.253** | 0.204* |
| | | | | (0.119) | (0.127) | (0.096) | (0.105) | (0.124) | (0.109) | (0.105) |
| Unemployment Rate (%) | -0.060** | -0.005 | -0.070 | 0.006 | -0.008 | -0.074*** | -0.060** | 0.033 | -0.056** | -0.044 |
| | (0.024) | (0.027) | (0.053) | (0.027) | (0.025) | (0.027) | (0.024) | (0.032) | (0.024) | (0.032) |
| Unemp. X Neigh. Attributes | | | | -0.142*** | -0.158*** | 0.046 | -0.011 | -0.145*** | -0.031 | -0.030 |
| | | | | (0.028) | (0.030) | (0.029) | (0.034) | (0.030) | (0.028) | (0.031) |
| N | 15,733 | 3,208 | 4,856 | 15,733 | 15,733 | 15,733 | 15,733 | 15,733 | 15,733 | 15,733 |
| adj. R² | 0.131 | 0.182 | 0.070 | 0.135 | 0.135 | 0.132 | 0.131 | 0.135 | 0.131 | 0.131 |
| F | 10.4*** | 0.2 | 13.0*** | 11.1*** | 11.4*** | 8.7*** | 7.8*** | 9.9*** | 7.9*** | 8.2*** |
| Month-of-sale FE | Y | Y | Y | Y | Y | Y | Y | Y | Y | Y |
| Month-of-previous-sale FE | Y | Y | Y | Y | Y | Y | Y | Y | Y | Y |

Notes: Standard errors are reported in parenthesis. Across all specifications, we cluster standard errors across MODZCTAs. Panel A shows the results focused on the pandemic-induced effects during first wave of infections between March and July for each year from 2003 through 2020, whereas Panel B shows results focused on the second wave between August and December for each year from 2003 through 2020. The estimation samples underlying the results are restricted to repeatedly-sold non-investor owned one or two-family properties. Accordingly, all variables are given in changes between sales, i.e., the dependent variable is given by the change in sale prices between sales in $100,000. Column (1) repeats the baseline results from Table 3. Columns (2) and (3) show the effects for homes priced in the bottom and top quartiles of pre-pandemic sale prices. Columns (4) through (11) show case and unemployment effects and their interactions with various neighborhood attributes. The interaction variables are indicators that equal 1 if a home is sold in a neighborhood with a above median characteristic. The indicators for ``BBL Density'', for example, marks MODZCTAs with above median population per BBL. Indicators for ``Public Transit'', ``Long Commute'', ``Limited English'', ``College Educated'', and ``Foreign Born'' represent MODZCTAs with above median population shares for residents that, for example, use public transit, have a commute longer than 30 minutes, are college educated, etc. Statistical significance at the conventional 10%, 5%, and 1% thresholds is given by *, **, ***, respectively.

case rates. The estimates given in columns (9) and (10) lend support to this hypothesis and show that the *contagion* effect is roughly half for properties located in these neighborhoods.

Overall, these price effect dynamics across various neighborhood characteristics are similar during the first and second wave of the pandemic. Interestingly, the differences in *contagion* price effects are less pronounced during the second wave of the pandemic, while the differences in *income* effects intensify with in neighborhoods with above median density, public transit dependence and rental shares.

## 5.3 Robustness Analysis

We conduct a host of robustness analyses to test the sensitivity of our primary COVID-19 price effects (as initially reported in columns (3) and (7) of Table 3). We begin by adjusting the restrictions on our repeat-sales sample. The primary results are based on transactions of properties repeatedly sold either between March and July 2003-2020 (i.e., first infection wave) or between August and December 2003-2020 (i.e., second wave of infections). In column (1) of Table 5, we report the COVID-19 and unemployment price effects when we relax these restrictions and estimate the average point estimates for the full sample instead. The coefficient estimates on case and unemployment rates are consistent in magnitude and are statistically significant at 1% level.

Next, we consider the sensitivity of our result against the presence of influential outliers. One criticism against the use of repeat-sales samples has arisen from the literature of price indexes, which have been shown to be more sensitive to outliers when relying on a repeat-sales sample (see, for example, Wallace and Meese (1997)). Although we are not using the repeat-sales sample to construct a price index, we investigate the potential impact of outliers. Excluding transactions that are more or less than three standard deviations removed from the average change in sale prices between repeat sales reduces the number of observations by 187 transactions for the first-wave sample and 158 transactions for the second-wave sample. The coefficient estimates of interest remain stable and are statistically significant at the conventional levels (see column (2) of Table 5).

Another potential concern with repeat sales is potential changes in unobservable property characteristics between sales. While the PLUTO$^{TM}$ database indicates whether a home has been altered from the perspective of the assessed value for taxation purposes, and we control for this alteration, it is possible that some properties undergo changes that are not necessarily captured by the tax authority and therefore unobservable to us. This may be particularly likely for homes purchased for investment purposes, which may receive a few non-structural updates/renovations and are resold quickly. Although we attempt to exclude investor-owned properties based on property owner names, it is possible that some investment-driven purchases remain in the estimation samples. To avoid the potential influence of such properties, we take two distinct approaches to exclude them. In the first, we restrict the sample to properties for which the duration between sales exceeds three years. We believe that these homes are unlikely to be 'flipped' investment properties and therefore less likely to suffer from the potential bias of unobservable and valued home characteristics that have changed between sales. Column (3) of Table 5 shows that this removes a large number of transactions of repeat-sales homes. The point estimates on the effect of changes in cumulative case rates increase in magnitude, while statistical significance only slightly suffers from the reduction in observations. The unemployment effect estimate nearly doubles in size and becomes statistically significant for the first-wave sample, while the second-wave sample estimate remains qualitatively and quantitatively consistent. That is, among homes that are less frequently sold, the COVID-19 case and unemployment price discounts tend to be larger than for more frequently sold properties.

An alternative strategy to address the issue of more frequently sold investment properties is to the restrict the sample of homes in terms of the number of transactions per property. Limiting the number of property transactions to less than four per home reduces the number of observations of repeat-sales properties by around 1,700 and renders the pandemic-induced *contagion* and *income* effects on prices largely unchanged (see column (4) of Table 5).

A final concern is the possibility that we may confound the impact of the local intensity of the pandemic, measured by the cumulative MODZCTA Coronavirus case rate, with the effects of the shutdown of NYC. That is, after the total number of Coronavirus infections rose to 7,102 in NYC, Governor Cuomo signed the New York State on PAUSE executive order shutting down non-essential business, prohibiting gatherings and imposing social distancing by March 22, 2020. The shutdown lasted through June 8, 2020, at which point NYC went through a gradual reopening period. Since cases rose significantly during the first wave of the pandemic – after the shut down – it is possible that we falsely attribute the price effects of this executive shutdown order to the initial rise in the rates of infections. To address this concern, we integrate a control variable that indicates whether the repeat sale of a property

Table 5: Robustness Tests of Repeat-Sales COVID-19 Price Effects

|  | (1) | (2) | (3) | (4) | (5) | (6) | (7) | (8) |
|---|---|---|---|---|---|---|---|---|
|  | Sample Restrictions | | | | | Alternative Pandemic Measures | | |
|  | All Repeat Sales | No Out-liers | > 3 yrs. between Sales | # of Sales < 4 | NYC Shutdown | 30-Day Lag | 60-Day Lag | Borough Level Rates |
| **Panel A: First Wave of Infections (03/01-07/31, 2003-2020)** | | | | | | | | |
| Cases per 100,000 ('000) | -0.594*** | -0.669* | -0.893 | -0.703* | -0.719** | -0.631* | -0.422 | -1.005* |
|  | (0.211) | (0.351) | (1.856) | (0.359) | (0.358) | (0.344) | (0.394) | (0.580) |
| Unemployment Rate (%) | -0.053*** | -0.068*** | -0.074*** | -0.041 | -0.039 | -0.037 | -0.043* | -0.232* |
|  | (0.018) | (0.018) | (0.026) | (0.026) | (0.024) | (0.024) | (0.023) | (0.118) |
| NYC Shutdown |  |  |  |  | 0.238 |  |  |  |
|  |  |  |  |  | (0.443) |  |  |  |
| $N$ | 36,051 | 15,353 | 6,174 | 13,757 | 15,540 | 15,323 | 15,094 | 15,540 |
| adj. $R^2$ | 0.145 | 0.228 | 0.101 | 0.136 | 0.146 | 0.146 | 0.146 | 0.146 |
| F | 19.7*** | 21.4*** | 11.2*** | 15.0*** | 14.1*** | 14.3*** | 13.1*** | 13.8*** |
| **Panel B: Second Wave of Infections (08/01-12/31,2003-2020)** | | | | | | | | |
| Cases per 100,000 ('000) | -0.594*** | -0.481** | -1.060** | -0.633*** | - | -0.622** | -0.500 | -1.119*** |
|  | (0.211) | (0.198) | (0.441) | (0.242) | - | (0.297) | (0.318) | (0.344) |
| Unemployment Rate (%) | -0.053*** | -0.070*** | -0.064*** | -0.064*** | - | -0.067*** | -0.061*** | -0.166 |
|  | (0.018) | (0.019) | (0.024) | (0.024) | - | (0.022) | (0.021) | (0.105) |
| $N$ | 36,051 | 15,575 | 7,037 | 14,014 | - | 15,570 | 15,410 | 15,733 |
| adj. $R^2$ | 0.145 | 0.229 | 0.107 | 0.120 | - | 0.134 | 0.134 | 0.131 |
| F | 19.7*** | 21.5*** | 7.3*** | 10.1*** | - | 10.2*** | 9.5*** | 8.8*** |
| Home & Neigh. Charac. | Y | Y | Y | Y | Y | Y | Y | Y |
| Month-of-Sale FE | Y | Y | Y | Y | Y | Y | Y | Y |
| Month-of-Pre.-Sale FE | Y | Y | Y | Y | Y | Y | Y | Y |

Notes: Standard errors are reported in parenthesis. Across all specifications, we cluster standard errors across MODZCTAs. Panel A shows the results focused on the pandemic-induced effects during first wave of infections between March and July for each year from 2003 through 2020, whereas Panel B shows results focused on the second wave between August and December for each year from 2003 through 2020. Depending on the imposed sample restrictions, we observe sales in 135 to 155 MODZCTAs anywhere from as early as January 2003 through December 2020. Columns (1) through (4) show the COVID-19 cumulative case effect per 100,000 local MODZCTA residents (measured in 1,000 cases) as well as the unemployment effect on the change in sale prices between repeat transactions under varying sample restrictions. In column (5), we estimate the price effects of cumulative Coronavirus infections and unemployment while controlling for the signing of the New York State on PAUSE executive order. In columns (6) and (7), we report the 30-day and 60-day lagged effect of cumulative MODZCTA infection and unemployment rates. Finally, in column (8) we show the effect of changes in borough, rather than MODZCTA, cumulative infection and unemployment rates. Across all regressions, we control for time-varying home and neighborhood characteristics (i.e., whether a home was altered between sales and the share of investment-owned properties in a given neighborhood), month-of-sale and month-of-previous-sale fixed effects. Statistical significance at the conventional 10%, 5%, and 1% thresholds is given by *, **, ***, respectively.

occurred after much of the activities in NYC had been restricted. If the shutdown of NYC, rather than the intensity of the local outbreak, is the explanation for the estimated price effect, then the inclusion of this control variable should render our estimated case effect on home sale prices insignificant. If, however, we find that homes sold in neighborhoods with greater infection rates experience significantly larger price discounts even when controlling for the shutdown executive order, our estimate provides evidence of the *contagion* price effect (rather than the shutdown effect). Indeed, column (5) of Table 5 shows a statistically and economically significant price discount for homes located in neighborhoods with higher Coronavirus case rates, while the price effect of the shut-down is statistically insignificant.[18]

Combined, these robustness tests based on various sample restrictions produce convincing evidence in support of our primary estimates of the pandemic-induced price effects and the proper identification thereof. A final set of robustness analyses considers the timing of the evidenced pandemic-induced price discount and sensitivity of our results against more aggregated measures of the intensity of the COVID-19 pandemic and rise in unemployment. Similar to the sample restrictions, these tests produce estimates that reiterate and tend to support our primary findings. The results reported in columns (6) and (7) of Table 5, for example, show the estimated COVID-19 price discount with respect to 30- and 60-day lagged case and unemployment rates. While these lagged effects are quantitatively consistent with the primary results, the effects of 60-day lagged cumulative case rates are found to be statistically insignificant. Lastly, we consider case and unemployment rates at the more aggregated borough, rather than MODZCTA, level. Reassuringly, the estimates in column (8) of Table 5 show that our findings are generally robust against the extrapolation of MODZCTA case and unemployment rates.

## 6 Discussion and Potential Limitations

What is the economic significance of these estimates? Focusing on our repeat-sales sample results, our price estimate suggests that during the second wave of the pandemic the average sale price of a repeat-sale NYC home drops by close to $63,000 for every 1,000 novel Coronavirus infections added to the cumulative case rate per 100,000 residents. Between August and December, 2020 (the end of our sample), the change in MODZCTA case rates in NYC ranged anywhere from 100 to 7,100 and averaged 1,776 infections per 100,000 residents. Combining these statistics with our estimate suggests that the expected price effects may range from as little as $6,000 on the lower end to nearly $450,000 in neighborhoods with the largest increases in case rates and average around $100,000. A comparison of these figures to the average 2019 NYC sale price listed in Table 1 indicates that in the short-run COVID-19 may have reduced home values in NYC anywhere from 0.8% to over 50% depending on the severity of the local outbreak and 14% on average. Summary statistics for our repeat-sales sample suggest that for homes that sold at least twice between January, 2003 and December, 2019, the average NYC property appreciated by around $150,000 (around 27%) over the course of 4.5 years between sales. More recently, homes that sold at least twice between 2017 and 2019 appreciated by roughly $100,000 (around 20%) over the course of 3.7 years between sales. Accordingly, a back-of-the-envelope calculation based on our estimate indicates that during a five-month span in the latter half of 2020 COVID-19 eroded a value equivalent to an expected three- to four-year return on investment in residential one- or two-family home properties.

---

[18] We note that the gradual reopening was largely concluded by August 2020 and therefore we cannot estimate the effect of the NYC shut down during the second wave of the pandemic.

In our heterogeneity analysis, we find evidence of pandemic-induced compositional shifts. In other words, housing wealth appears to have accrued disproportionately to owners of higher valued houses during the first phase of the pandemic. Homeowners who are more affluent likely hold greater home equity, which can make them more mobile. Hence, they may be able to more easily avoid locations with higher risk of contagion. Conversely, lower income homeowners, who may have little home equity, likely cannot afford to be as risk averse.

While our empirical analyses produce consistent results that are in line with some of the early anecdotal evidence reported for NYC, we point out a few caveats regarding our findings. First, we note some data limitations. While our NYC real-estate data possess numerous property characteristics, neither the PLUTO$^{TM}$ or NYC Department of Finance data provide some of the detailed home characteristics typically associated with a home's value, such as the number of bedrooms or number of bathrooms. In absence of these attributes, it is possible that a non-repeat sales hedonic analysis could be imprecisely tuned to the nuance of each individual property.[19] The repeat-sales analysis, however, differences out time-invariant unobservable characteristics and should render unbiased coefficient estimates.[20] We also control for properties that have undergone major renovations, which abuts a major potential pitfall of the repeat-sales approach and further validates its benefits in our context.

---

[19] Similarly, apartment-level property characteristics are not available since the data are reported on the borough-block-lot level. In other words, two apartment units in the same building receive the property characteristics of the overall building, rather than the specific details of each unit. Since we focus on residential one- or two-family properties, this does not impact our analysis. And, while the majority of NYC real estate transactions indeed involve one- or two-family properties, it is noteworthy that this is not the case for the Manhattan residential real estate market, where the bulk of transactions involves apartments and condominiums. Consequently, the omission of apartments from our analysis implies that our results related to Manhattan likely comprise only a very small subset of the Manhattan entire real-estate market. Though, as many NYC residents will remind those living in Manhattan, NYC has five boroughs and thus our analysis can still offer a meaningful first glance at the impact of COVID-19 on this major U.S. urban center.

[20] It is possible that the alteration of a property may change the number of bedrooms and bathrooms. If such alterations are officially permitted and change the assessed value of a home, the PLUTO$^{TM}$ dataset reports such changes and we explicitly control for them via an indicator variable. If such alterations are unreported, even the repeat-sales estimator misses the changes in home attributes and it is one weakness of this analysis that we attempt to address via a few robustness tests reported in Section 5.

Finally, there may be some limits to the external validity of our study. First, our analyses pertain to NYC. While this is a key real estate market in United States and our results can produce a useful benchmark that may be indicative of the COVID-19 impact to be expected in other larger urban centers, it may not be reflective of all U.S. real estate markets and their responses to the current pandemic. Moreover, our analyses apply to one- and two-family homes. The estimated pandemic impact on these types of residential homes may be different from that of other residential properties, such as apartments and condominiums, or commercial real estate more generally, and leaves room for future investigations that are beyond the scope of this paper.

# 7 Conclusion

In this paper, we model how individual residential real estate sale prices in NYC were related to growing numbers of cases of the novel Coronavirus in nearby hospitals during the early months of the pandemic. We also consider how changes in employment across neighborhoods factor into the hedonic modeling. The novel Coronavirus pandemic has been rampant and relentless throughout the US, and it has reinvented the way most people live, with many working from their homes. While safe and effective vaccines have been available, there are several uncertainties as to whether this virus will completely disappear any time in the foreseeable future. This leads one to ponder: will the novel Coronavirus continue to mutate and linger around for some time, and how will this impact residential real estate markets?

Even if the novel Coronavirus does not persist for as long as some other viruses have,[21] there are likely to be additional waves (as has been the case with the Delta and Omicron variants in 2021-22) before the virus is completely eradicated. These ongoing concerns could lead to structural change in how people think about their desired dwelling locations. Density in big cities in the U.S. can facilitate transmission, due to close proximity of other individuals in tight living quarters, use of mass transit due to scarce parking and road traffic, crowded public schools, and houses that are closer together than those in the suburbs. Even within many large U.S. cities, neighborhoods with lower-valued real estate often have residences that are closer together. Therefore, these concentrated neighborhoods may be more prone to spreading the virus than higher-valued homes in more affluent neighborhoods with larger lots of land and plentiful parking available for private vehicles. Larger homes in the suburbs facilitate home offices that may be precluded in large cities by smaller residences. Neighborhoods with more cases may be more adversely affected and therefore residents may be more likely to choose to move out, which may adversely affect residential real estate prices. These anecdotes underscore the importance of studying the relationships between residential real estate prices and the incidence of the novel Coronavirus. The demographics and real estate markets of NYC facilitate such an analysis for residential real estate sales across the five boroughs.

---

[21] One dismal tidbit of history is that the 1918 Spanish Flu pandemic was still around in 2010 in the form of the H1N1 flu.

Our results have shed new light on these issues. In a traditional hedonic specification, where we include the rate of cumulative cases as a regressor, we find a negative statistically significant relationship between case numbers and one- and two-family residential real estate prices across all neighborhoods within a NYC borough. This initial analysis raises a question: does COVID-19 have an impact on prices, or is the estimated average price effect at the neighborhood level the result of a change in the local composition of sold properties?

We drill down deeper to answer this question. Our repeat-sales approach enables us to identify the relationships between infections (i.e., through the change in nearby cumulative cases between the two sales dates of a given property) and sale prices (i.e., the change in sales price for a property's two sales) at the property rather than average neighborhood sale price level. Controlling for observable and unobservable time-invariant home and neighborhood characteristics, we find that a given property's sales price falls by an average of approximately $100,000 in response to an additional 1,000 Coronavirus infections per 100,000 residents of the local MODZCTA. Up through December of 2020, the MODZCTAs with the least number of infections relative to the local population experienced slightly less than a 1% decrease in property prices due to COVID-19, while the MODZCTAs with the greatest numbers of infections per resident saw prices fall by nearly half. This evidence of heterogeneity across different value homes implies that the most highly valued homes lost relatively less value than the least valued homes. This wealth disparity may have been facilitated by the out-migration of wealthier homeowners to the highly competitive housing markets in outlying areas, which may have been perceived as safer from the risks of COVID-19 in NYC. As suburban dwelling demand increased to the point homes were seeing multiple offers above asking prices, owners of lower-valued homes may have been precluded from leaving the City.